\documentclass[mathleft
]{an}
\usepackage{graphicx}
\usepackage{epsfig}
\usepackage{times}
\begin{document}


\title{Photometric amplitudes and phases of B-type main sequence pulsators:
sources of inaccuracy}

\author{Wojciech Szewczuk\thanks{\email{szewczuk@astro.uni.wroc.pl}}
\and  Jadwiga Daszy\'nska-Daszkiewicz\thanks{
  \email{daszynska@astro.uni.wroc.pl}\newline}
}
\titlerunning{Photometric observables of B-type pulsators}
\authorrunning{ W. Szewczuk \& J. Daszy\'nska-Daszkiewicz}
\institute{
Instytut Astronomiczny, Uniwersytet Wroc{\l}awski, Poland}

\received{}
\accepted{}
\publonline{later}


\abstract{%
  We discuss all possible sources of uncertainties in theoretical values
  of the photometric amplitudes and phases of B-type main sequence pulsators.
  These observables are of particular importance because they contain
  information about the mode geometry as well as about stellar physics.
  Here, we study effects of various parameters coming both from theory
  of linear nonadiabatic oscillations and from models of stellar atmospheres.
  In particular, we show effects of chemical composition, opacities and,
  for the first time, effects of the NLTE atmospheres.}

\maketitle

\section{Introduction}
To construct a seismic model of a star, knowledge about the geometry of observed modes
is a precondition. In the case of B-type pulsators, mode identification cannot be done directly
from oscillation spectra because they are sparse and lack equidistant patterns.
An alternative way is to make use of the fact that information about the mode degree, $\ell$,
and the azimuthal order, $m$, is embedded in the photometric and spectroscopic variations
of a pulsating star. Ones of the most popular tools to identify a pulsation mode
are the amplitude ratios and phase differences in various photometric passbands.
In the case of zero-rotation approximation, these observables are independent
of the azimuthal order, $m$, and inclination angle, $i$.

The semi-analytical expression for the bolometric light variation
was formulated by Dziembowski \cite {dziemb}.
Balona \& Stobie \cite{balst} and Stamford \& Watson \cite{stwa} expanded this expression for the light
variation in photometric passbands. They showed that modes with
different values of $\ell$ are located in separated parts on the amplitude ratio
$vs.$ phase difference diagrams based on multicolour photometry.
Subsequently, this method was applied to various types of pulsating stars
by Watson \cite{watson}. Cugier, Dziembowski \& Pamyatnykh \cite{cdp} improved the method by including nonadiabatic effects
in calculations for the $\beta$ Cephei stars.
Effects of rotation on photometric observables were studied by Daszy\'nska-Daszkiewicz et al.\,(\cite{ref3})
for close frequency modes and by Townsend \cite{townsend2003} and Daszy\'nska-Daszkiewicz, Dziembowski \& Pamyatnykh \cite{dd2007}
for long-period g-modes.

The goal of this paper is to examine all possible effects on
theoretical values of the photometric amplitude ratios and phase differences
for early B-type pulsators. As an example, we consider the main sequence models
with a mass of 10 $M_\odot$ and low degree modes, $\ell$=0, 1, 2.
All effects of rotation on pulsation are neglected.
In Section 2, we recall basic formulas and describe our models.
Effects of parameters coming from linear non\-adiabatic theory of
stellar pulsation are presented in Section 3. Effects
of atmospheric parameters are discussed in Section 4.
The last section contains Conclusions.

\section{Pulsational changes of a star's brightness}
Stellar pulsations cause the changes of temperature, normal to the surface element
and pressure. If all effects of rotation on pulsation can be ignored,
then the total amplitude of the light variation in the passband $\lambda$
can be written in the following complex form (Daszy\'nska-Daszkiewicz et al.\,\cite{ref3}):
\begin{equation}
  \label{eq1}
  \mathcal{A}_{\lambda}(i)=-1.086 \varepsilon Y_{\ell}^m(i,0)b_{\ell}^{\lambda}(D_{1,\ell}^{\lambda} f+D_{2,\ell}+D_{3,\ell}^{\lambda}),
\end{equation}
where $\varepsilon$ is the intrinsic mode amplitude, $Y_{\ell}^m$ -- the spherical harmonic
and $i$ -- the inclination angle.
The $D_{1,\ell}^{\lambda}\cdot f$ product stands for temperature changes, where
\begin{equation}
  \label{eq2}
  D_{1,\ell}^{\lambda}=\frac{1}{4} \frac{\partial \log(\mathcal{F}_{\lambda}|b_{\ell}^{\lambda}|)}
  {\partial \log T_{\rm eff}},
\end{equation}
and $f$ is the nonadiabatic complex parameter describing the amplitude of the radiative
flux perturbation to the radial displacement at the photosphere level
\begin{equation}
\label{f_par}
\frac{ \delta {\cal F}_{\rm bol} } { {\cal F}_{\rm bol} }=
{\rm Re}\{ \varepsilon f Y_\ell^m(\theta,\varphi) {\rm e}^{-{\rm i}
\omega t} \}.
\end{equation}
Geometrical term, $D_{2,\ell}$, is given by
\begin{equation}
  \label{eq3}
  D_{2,\ell}=(2+\ell)(1-\ell),
\end{equation}
and the pressure term, $D_{3,\ell}^{\lambda}$, by
\begin{equation}
  \label{eq4}
  D_{3,\ell}^{\lambda}=-\left( 2+ \frac{\omega ^2R^3}{GM} \right) \frac{\partial \log(\mathcal{F}_{\lambda}|b_{\ell}^{\lambda}|)}
  {\partial \log g_{\rm eff}^0}.
\end{equation}
${\cal F}_\lambda$ is the flux in the passband $\lambda$ and $b_{\ell}^{\lambda}$
is the disc averaging factor defined by
\begin{equation}
  \label{eq5}
  b_{\ell}^{\lambda}= \int _0^1 h_{\lambda} (\mu) \mu P_\ell(\mu) d\mu
\end{equation}
where $h_{\lambda} (\mu)$ is the limb darkening law and $P_\ell$ is the Legendre polynomial.
Other parameters have their usual meaning.

From the above expressions, one can see that two inputs are needed to compute
theoretical values of the photometric amplitudes and phases.
The first input comes from the nonadiabatic theory of stellar pulsation
and this is the $f$-parameter (Eq.\,\ref{eq1} and \ref{f_par}).
The second input is derived from models of stellar atmospheres and these
are the flux derivatives over effective temperature and gravity (Eq.\,\ref{eq2} and \ref{eq4}),
as well as limb-darkening and its derivatives (Eq.\,\ref{eq2}, \ref{eq4} and \ref{eq5}).

All computations were performed using the Warsaw-New Jersey evolutionary code
and the linear nonadiabatic pulsation code of Dziembowski (1977).
We considered \linebreak opacity tables from OPAL (Iglesias \& Rogers \cite{OPAL}) and OP (Seaton \cite{OP}) projects,
and two determinations of the solar chemical composition:
GN93 (Grevesse \& Noels \cite{GN93}) and A04 (Asplund et al.\,\cite{A04}).
As for models of stellar atmospheres, we used Kurucz models (Kurucz \cite{ref4}), computed
within the LTE approximation, and TLUSTY models (Lanz \& Hubeny \cite{ref6}) which include the NLTE effects.
Here, we adopted nonlinear limb darkening law as defined by Claret \cite {claret2000}
\begin{equation}
h_\lambda(\mu)=2~\frac{1 -\sum\limits_{k=1}^4 a_k^\lambda (1
-\mu^{k/2})} {1-\sum\limits_{k=1}^4 \frac{k}{k+4} a_k^\lambda }.
\end{equation}
In the case of Kurucz models, we relied on the limb darkening coefficients of Claret \cite {claret2000}.
For TLUSTY models, we determined these coefficients by ourselves
and these results will be published elsewhere (Daszy\'nska-Dasz\-kie\-wicz \& Szewczuk \cite{ddsz2010}).

In all comparisons, we used a reference model computed with:
the OP opacity tables, the A04 mixture, hydrogen abundance of $X$=0.7,
metal abundance of $Z$=0.02, without overshooting from a convective core, $\alpha_{ov}$=0,
Kurucz models of stellar atmospheres with the metallicity of [m/H]=0.0 and
the microturbulent velocity of $\xi_t$=2 km/s.

\section{Uncertainties from the pulsation theory}
In the case of the B-type pulsating stars, values of the $f$-parameter
are mostly sensitive to hydrogen and metal abundances,
chemical mixture and hence opacities.

In Fig.\,1, we show effects of the hydrogen abundance, $X$, and metallicity, $Z$,
on photometric observables for the 10 $M_\odot$ model in the course of its main sequence evolution.
We considered the Str\"{o}mgren $uy$ passbands and three first radial modes: p$_1$, p$_2$, p$_3$.
In the left panel, we show the amplitude ratio, $A_u/A_y$,
and in the right one the corresponding phase difference, $\phi_u-\phi_y$,
as a function of $T_{\rm eff}$.
\begin{figure*}[!htb]
\includegraphics[width=58mm, angle=-90]{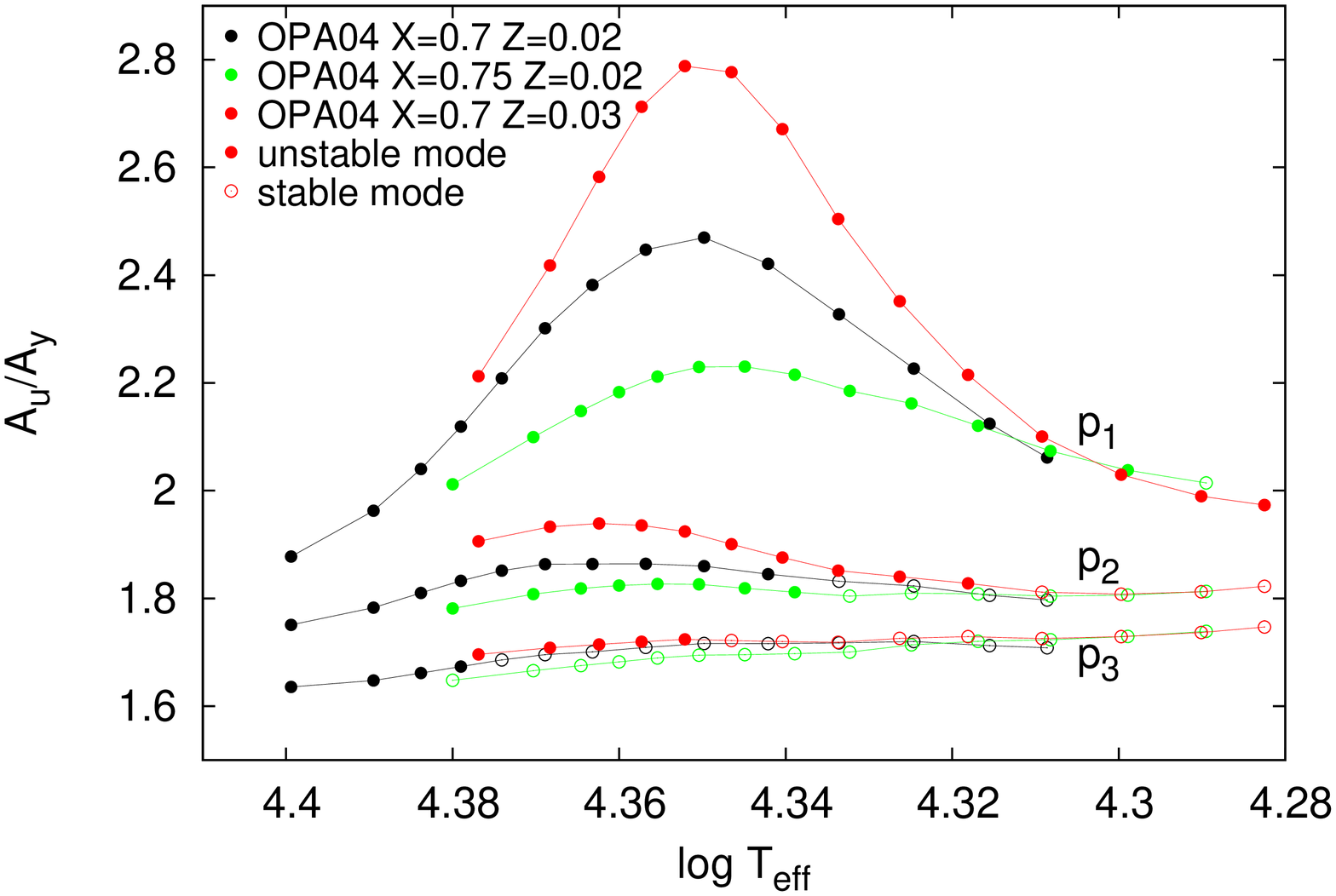}
\includegraphics[width=58mm, angle=-90]{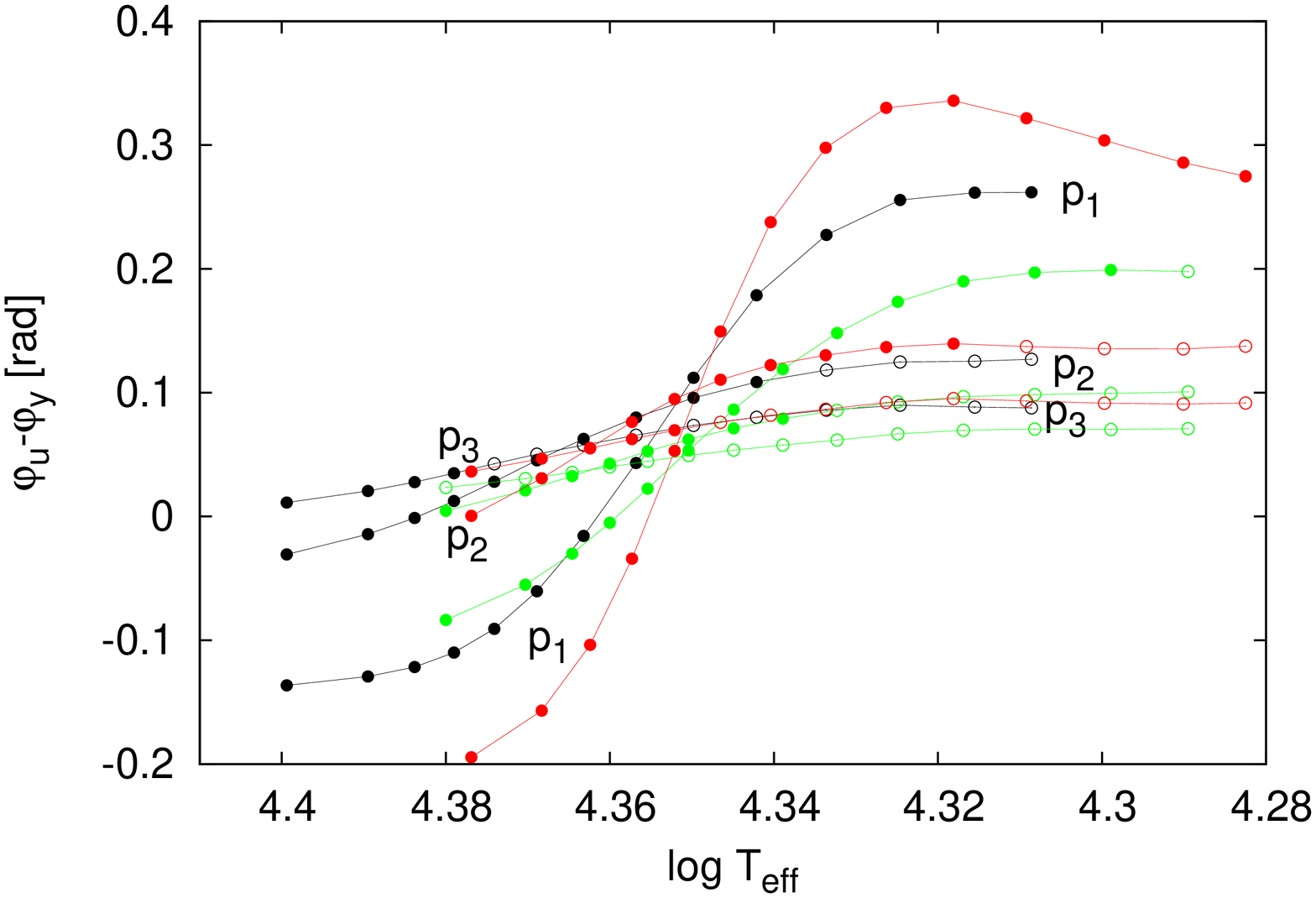}
\caption{Effect of the hydrogen ($X$) and metal ($Z$) abundance on photometric observables
for the three radial modes as a function of the effective temperature for the 10 $M_\odot$
main sequence model. In the left panel, we plot the amplitude ratios
in the Str\"omgren $uy$ filters and in the right one the corresponding phase differences.}
\label{f_amp_ZXl0}
\end{figure*}
We show computations obtained with $Z$=0.02 $vs.$ $Z$=0.03 and $X$=0.7 $vs.$ $X$=0.75.
As we can see, the amplitude ratios computed with $Z$=0.03 are larger than
those obtained with $Z$=0.02, whereas increasing hydrogen abundance, X, decreases
the amplitude ratios. Photometric observables of the radial fundamental mode, p$_1$,
are most sensitive to the chemical abundance.
\begin{figure*}[!htb]
\includegraphics[width=58mm, angle=-90]{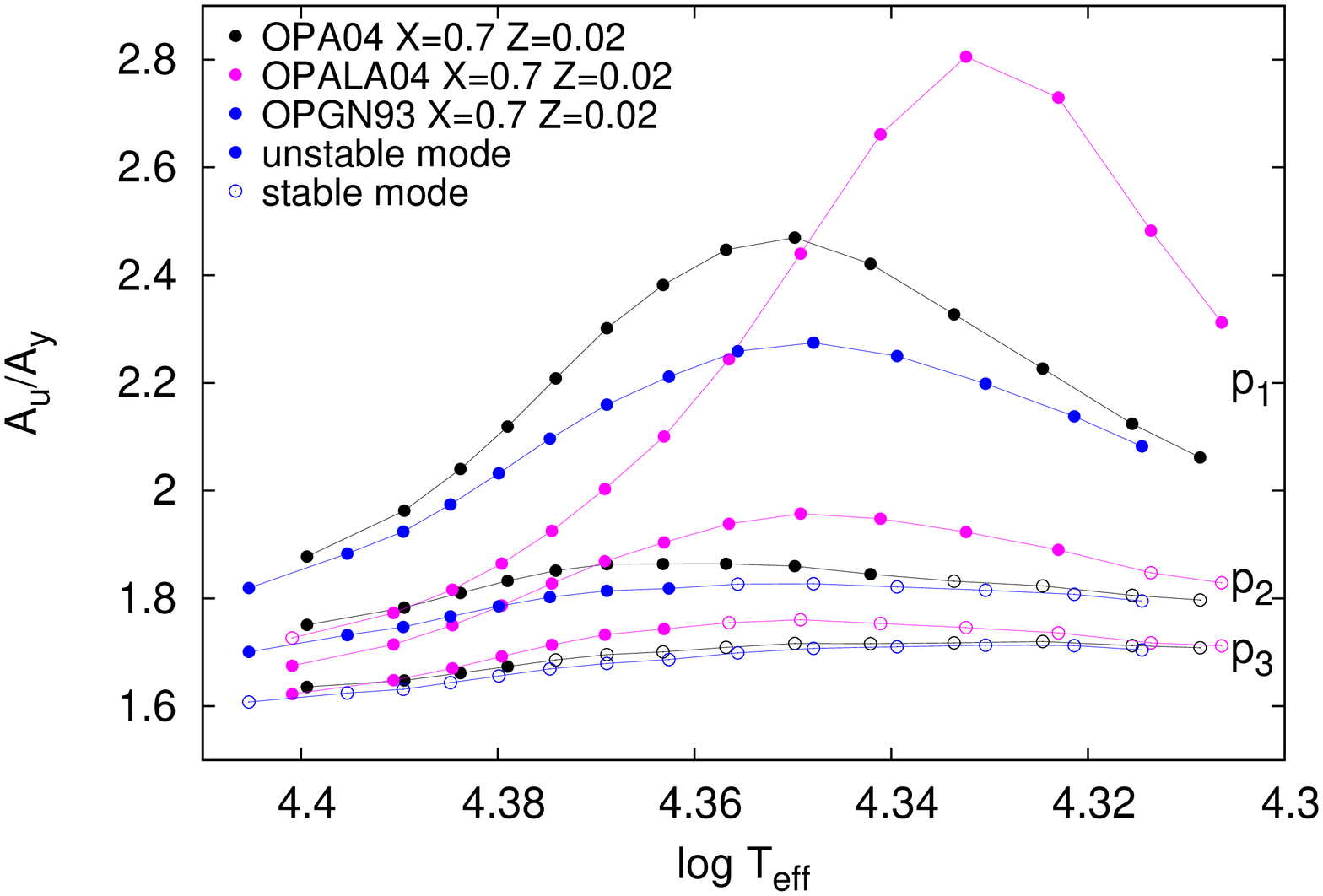}
\includegraphics[width=58mm, angle=-90]{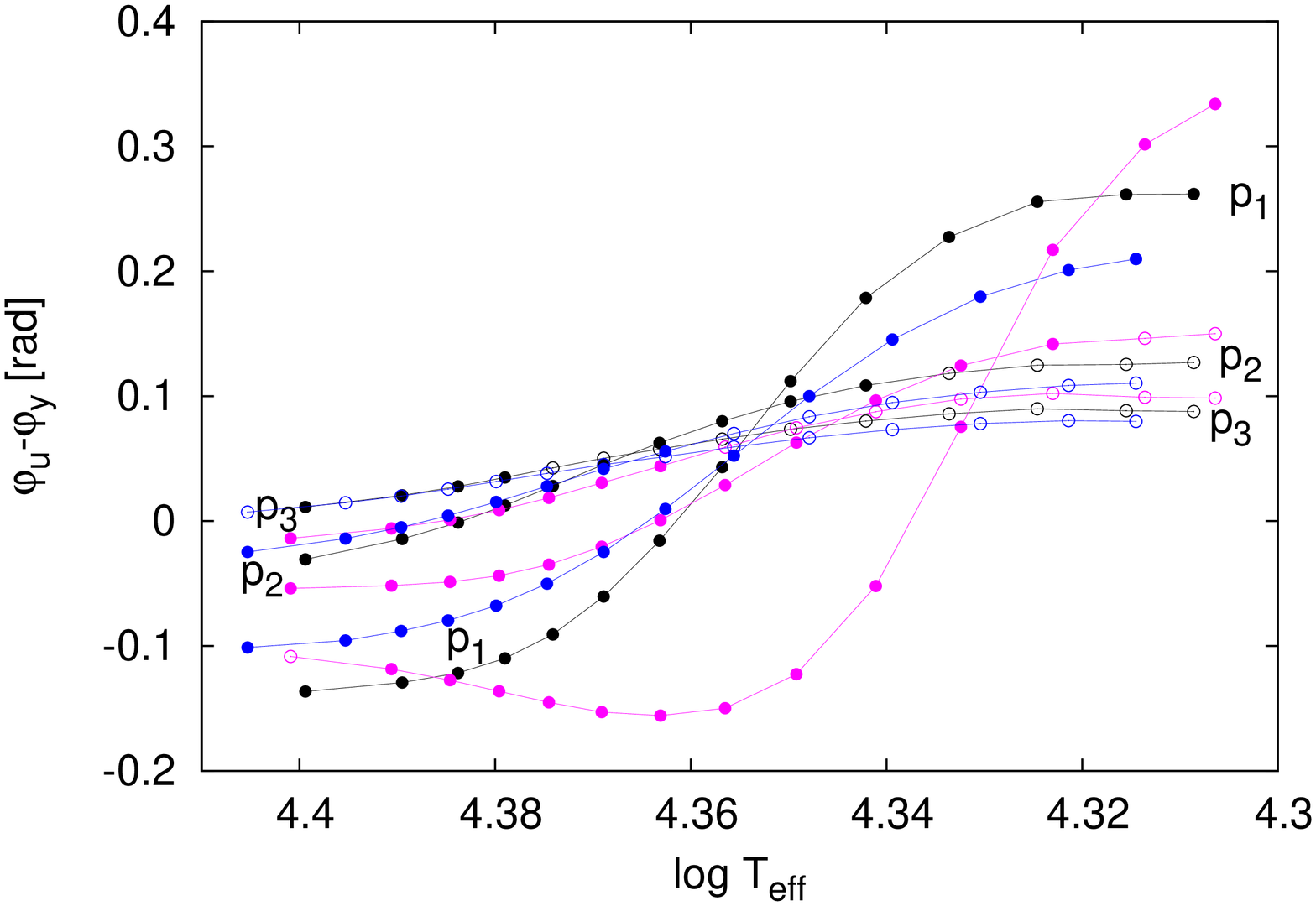}
\caption{The same as in Fig.\,\ref{f_amp_ZXl0} but effects of opacities and
         chemical mixture are presented.}
\label{f_amp_niepmixtl0}
\end{figure*}

Effects of the opacity tables and  chemical mixture for the radial modes are shown in Fig.\,\ref{f_amp_niepmixtl0}.
In this case, we compare photometric observables obtained with the OPAL $vs.$ OP tables
and the GN93 $vs.$ A04 mixture. Again, the largest effects are for the p$_1$ mode.
The amplitude ratios computed with the GN93 mixture are smaller than those obtained with A04,
because of relatively higher abundance of iron in A04.

Allowing overshooting from a convective core does not affect significantly photometric observables.

Computations for modes with $\ell$=1, 2 showed that the above discussed
parameters have much smaller influence on their photometric observables.

\section{Uncertainties from the atmosphere models}
The most important atmospheric parameters which can affect the photometric
amplitudes and phases of a pulsating star are: the metallicity parameter, [m/H],
the microturbulent velocity, $\xi_t$, and effects of NLTE.

In Fig.\,3, we show effects of [m/H] and $\xi_t$.
Here, we considered [m/H]=+0.5 and $\xi_t$=8 km/s, in addition
to our standard values: [m/H]=0.0 and $\xi_t$=2 km/s.
In general, effects of those parameters on photometric observables
are smaller than effects discussed in the previous section.

\begin{figure*}[!htb]
\includegraphics[width=58mm, angle=-90]{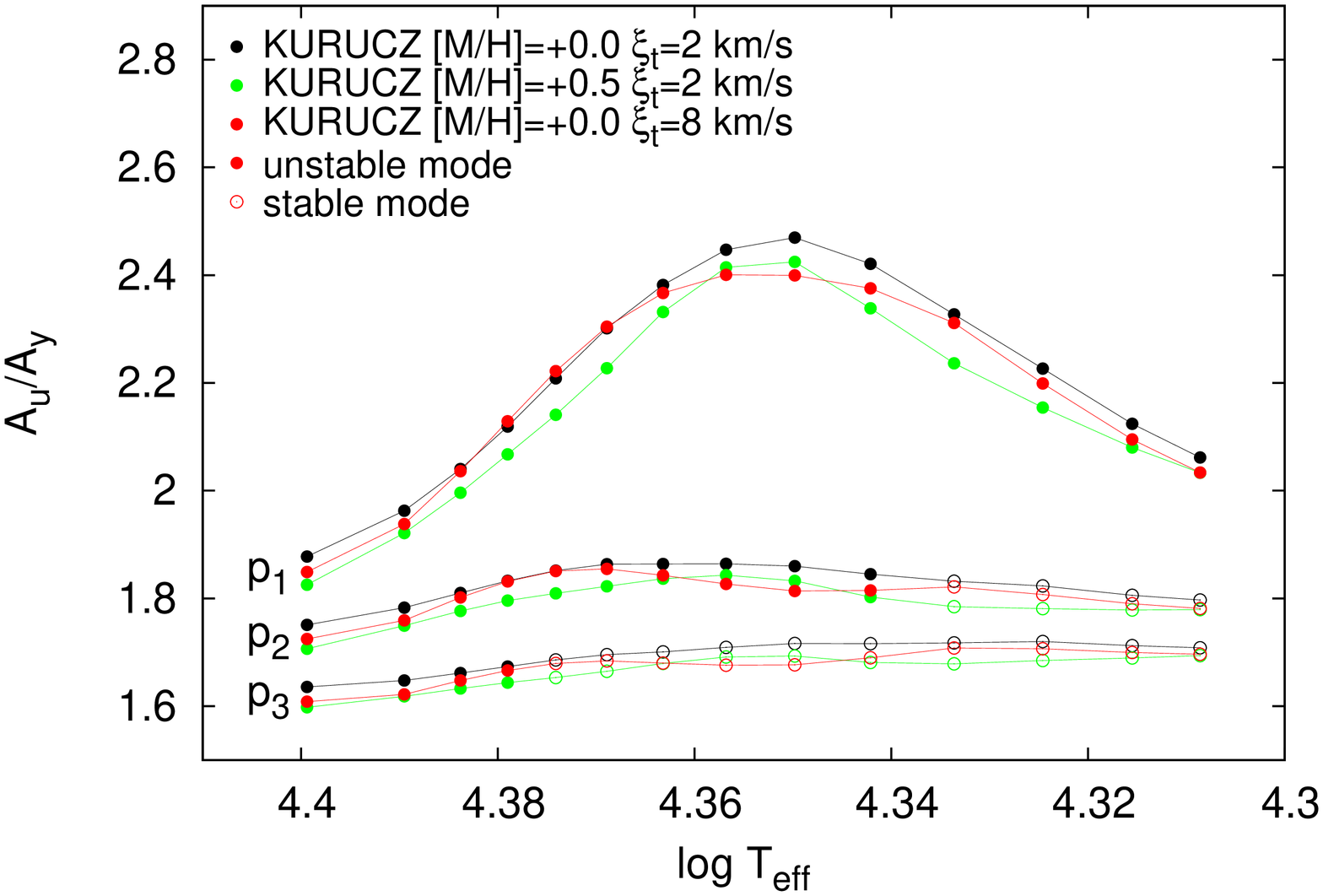}
\includegraphics[width=58mm, angle=-90]{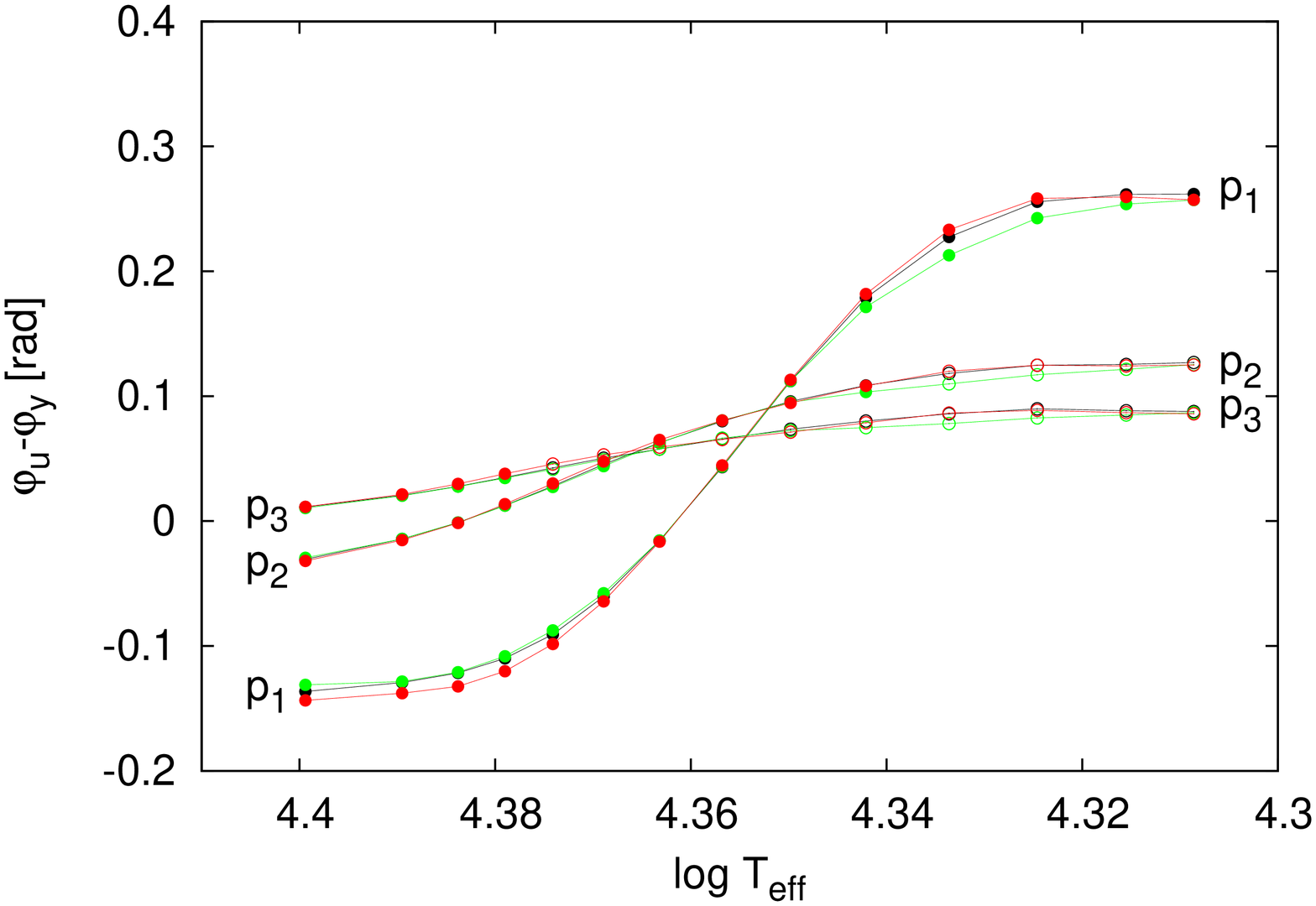}
\caption{The same as in Fig.\,\ref{f_amp_ZXl0} but effects of
         the atmospheric metallicity, [m/H], and the microturbulent velocity, $\xi_t$, are presented.}
\label{atm_amp_mettur}
\end{figure*}

\begin{figure*}[!htb]
\includegraphics[width=58mm, angle=-90]{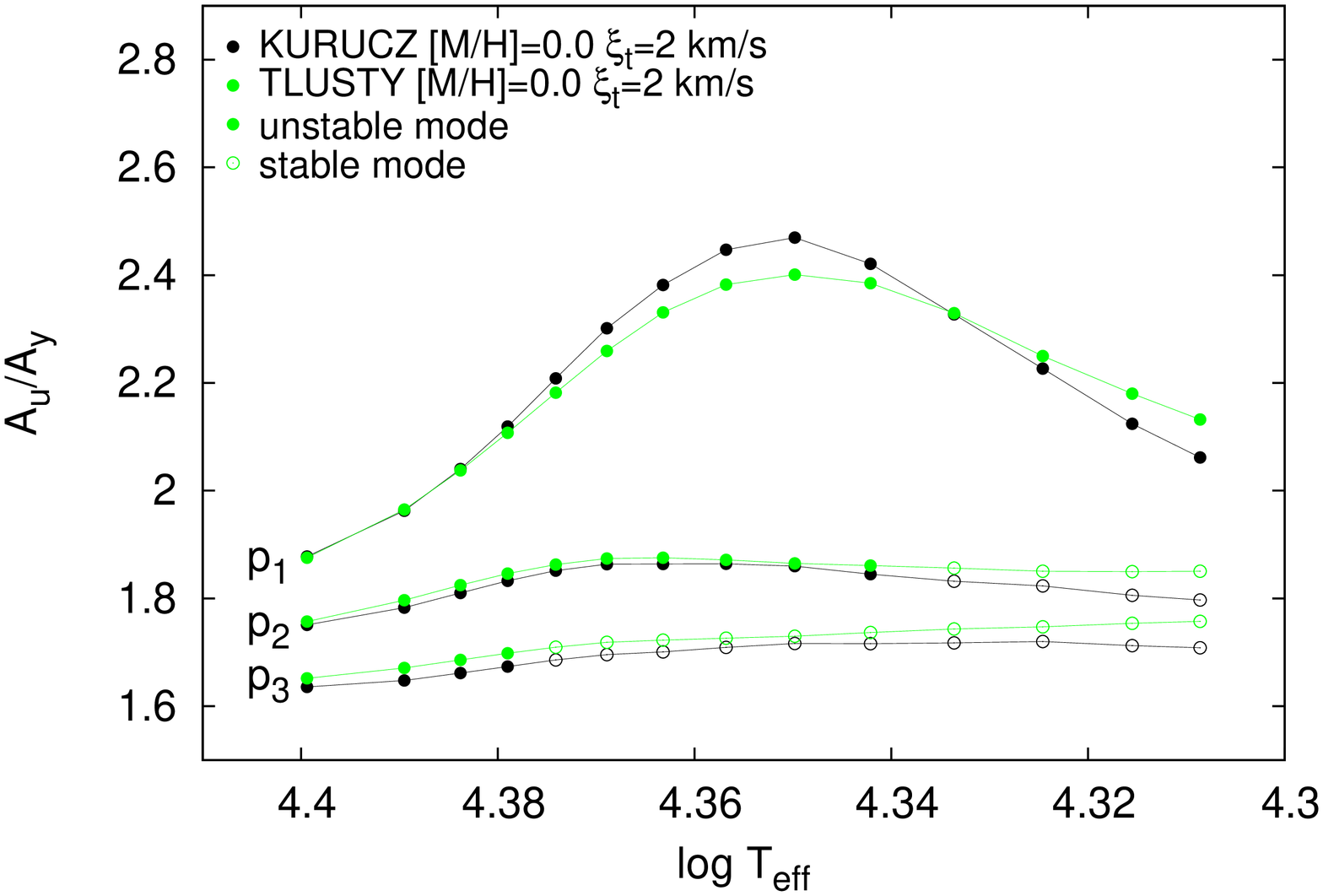}
\includegraphics[width=58mm, angle=-90]{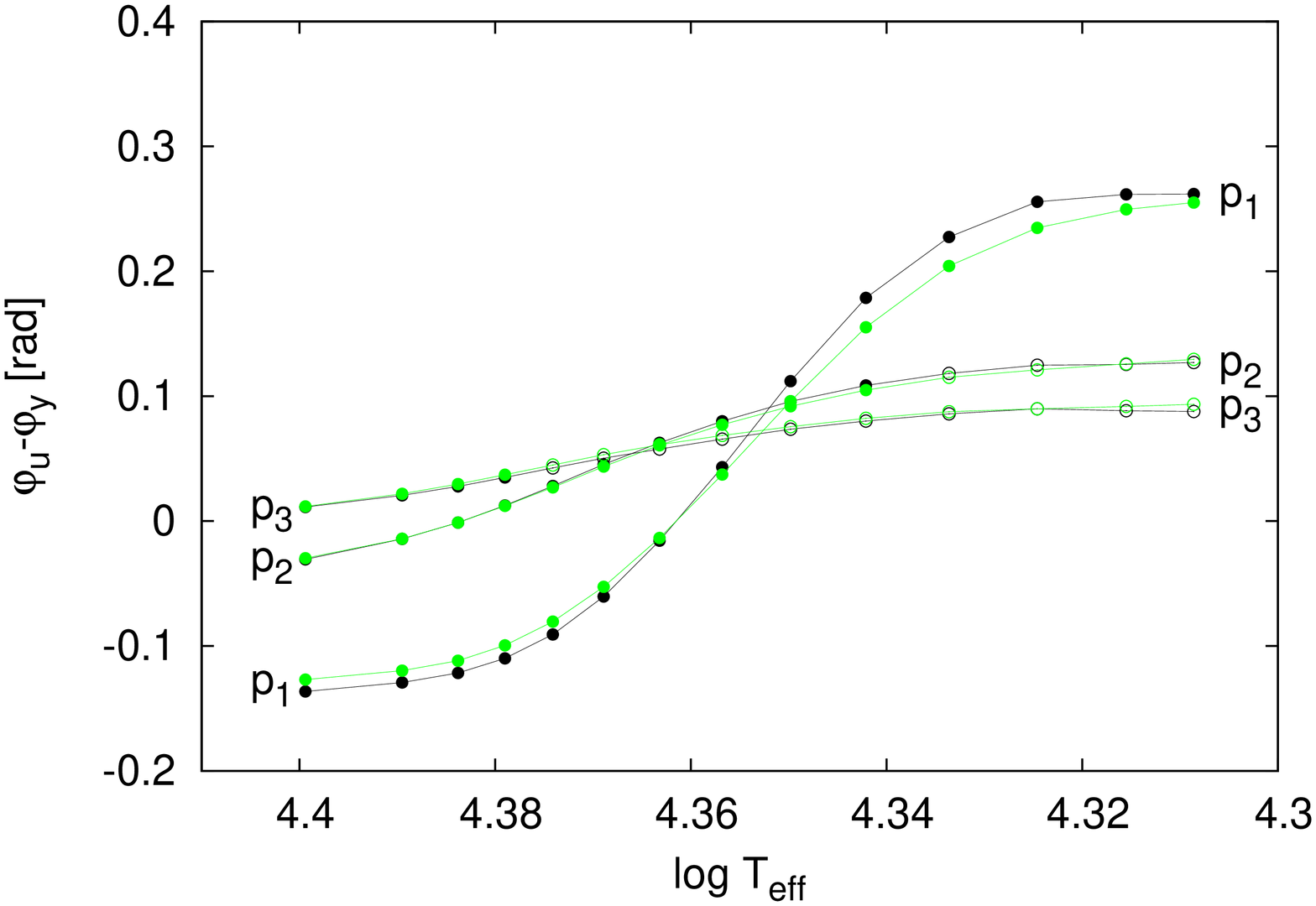}
\caption{Effect of NLTE on photometric observables
for the three radial modes as a function of the effective temperature for the 10 $M_\odot$
main sequence model. In the left panel, we plot the amplitude ratios
in the Str\"omgren $uy$ filters and in the right one the corresponding phase differences.}
\label{atm_amp_NLTEl0}
\end{figure*}

Subsequently, we studied effect of the assumption \linebreak of LTE in stellar atmosphere models.
In Fig.\,\ref{atm_amp_NLTEl0}, we compare computations with Kurucz LTE models
and TLUSTY NLTE models, at the same values of [m/H] and $\xi_t$=2 km/s,
for the radial modes. As we can see, the NLTE effects on photometric observables are
relatively small and comparable to effects of the atmospheric metallicity, [m/H],
or the microturbulent velocity, $\xi_t$ (cf. Fig.\,\ref{atm_amp_mettur}).

\begin{figure*}[!htb]
\includegraphics[width=58mm, angle=-90]{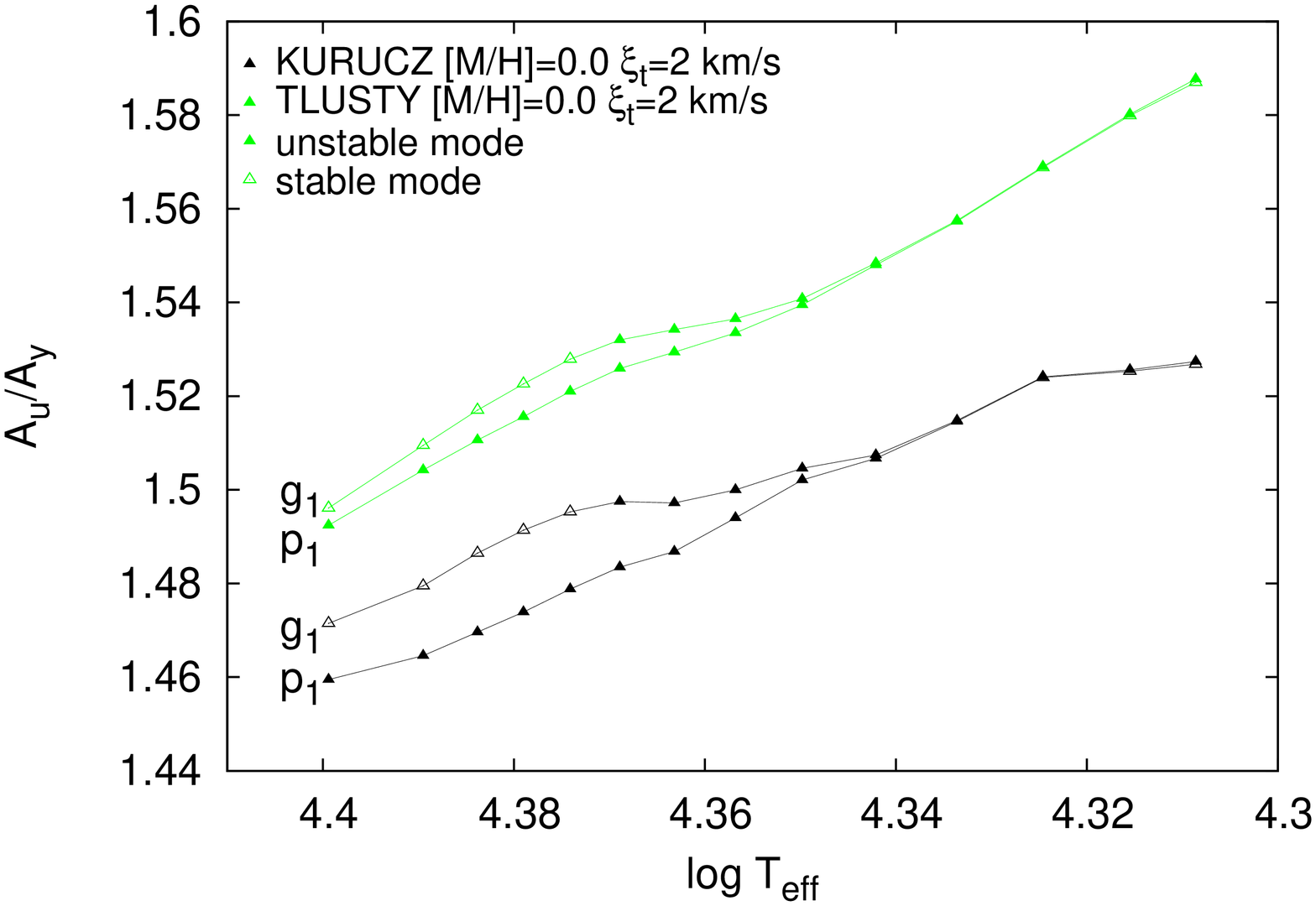}
\includegraphics[width=58mm, angle=-90]{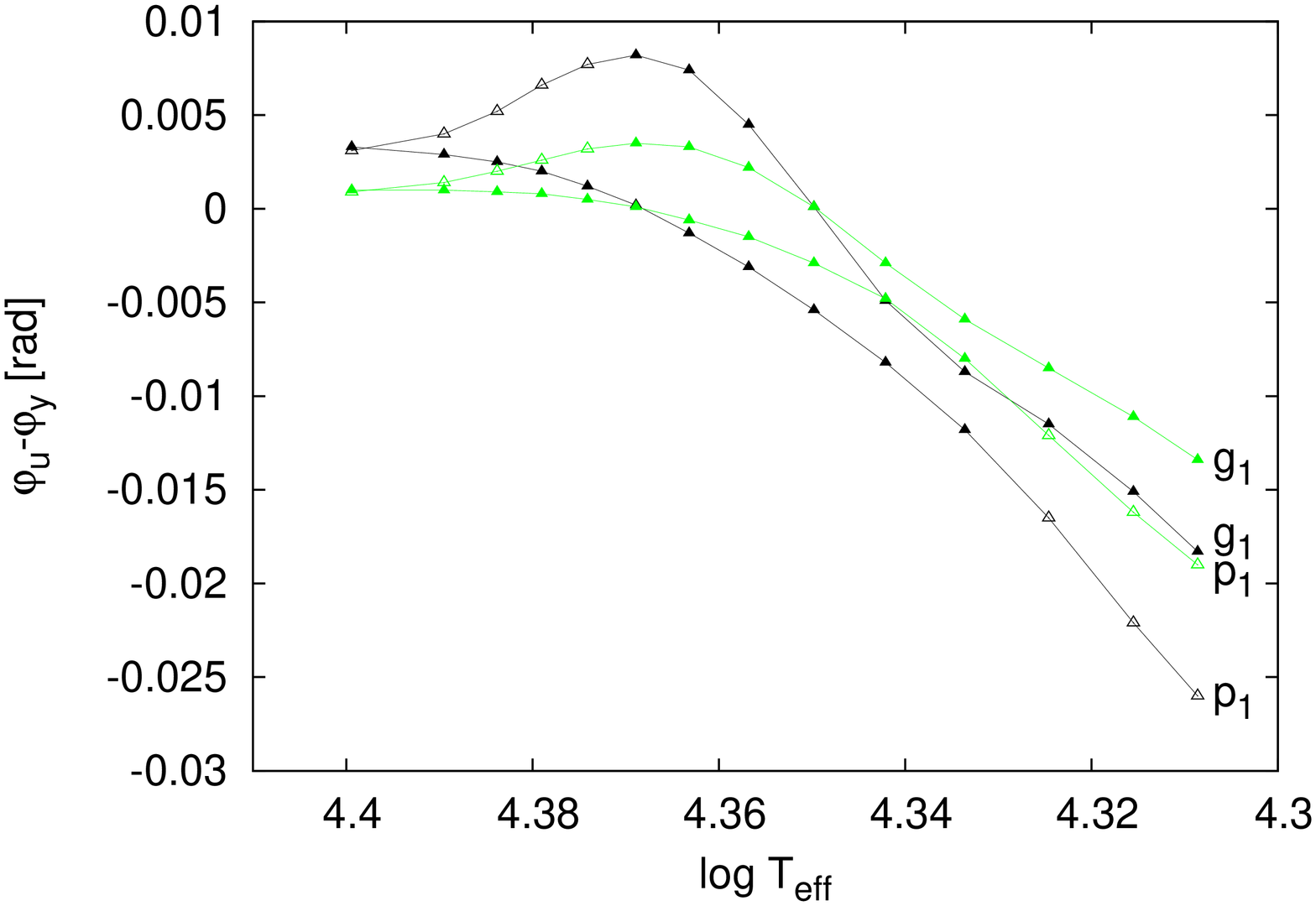}
\caption{The same as in Fig.\,\ref{atm_amp_NLTEl0} but for the two $\ell$=1 modes: p$_1$ and g$_1$. Note different scale on the Y axis.}
\label{aaaa}
\end{figure*}

\begin{figure*}[!htb]
\includegraphics[width=58mm, angle=-90]{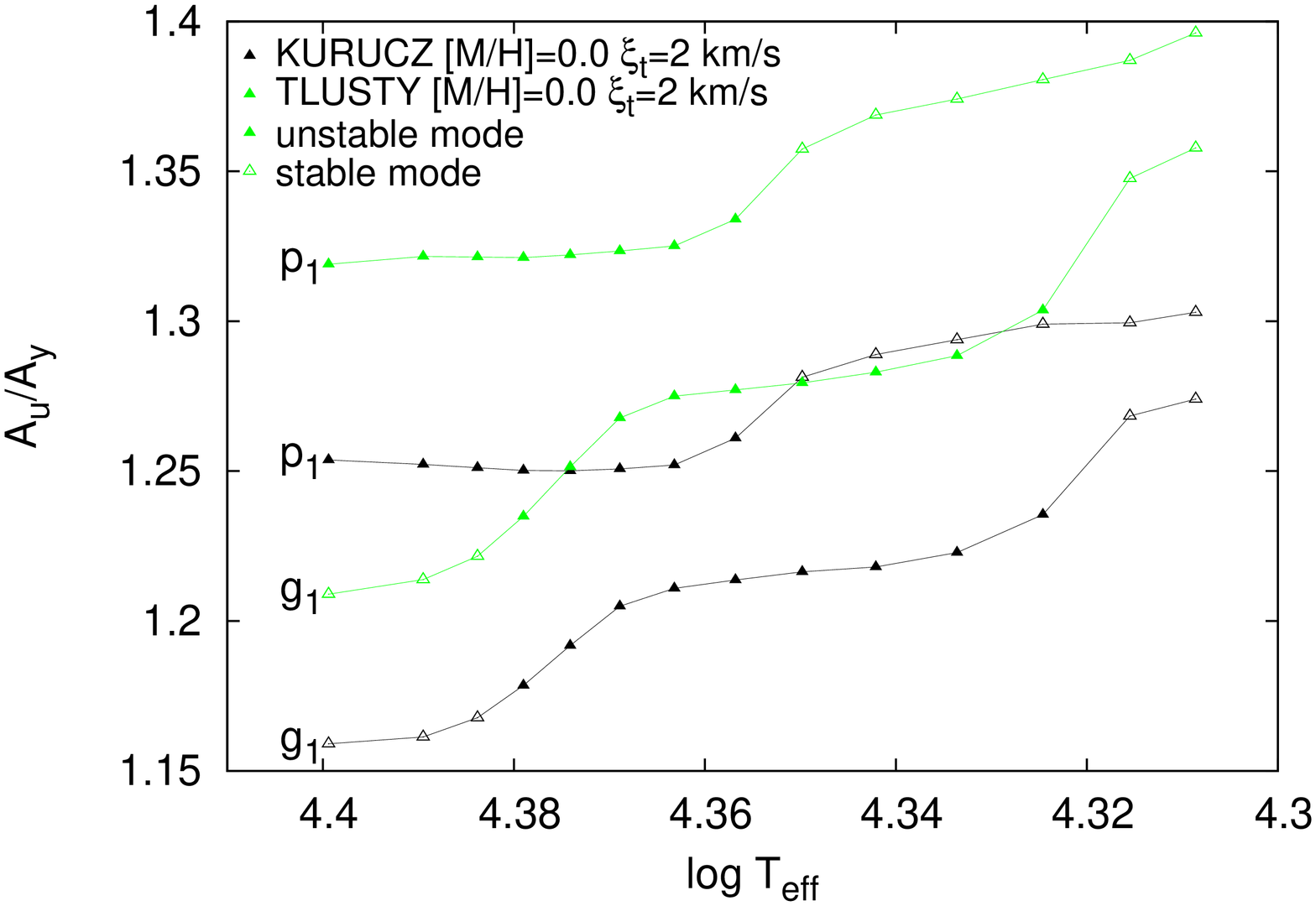}
\includegraphics[width=58mm, angle=-90]{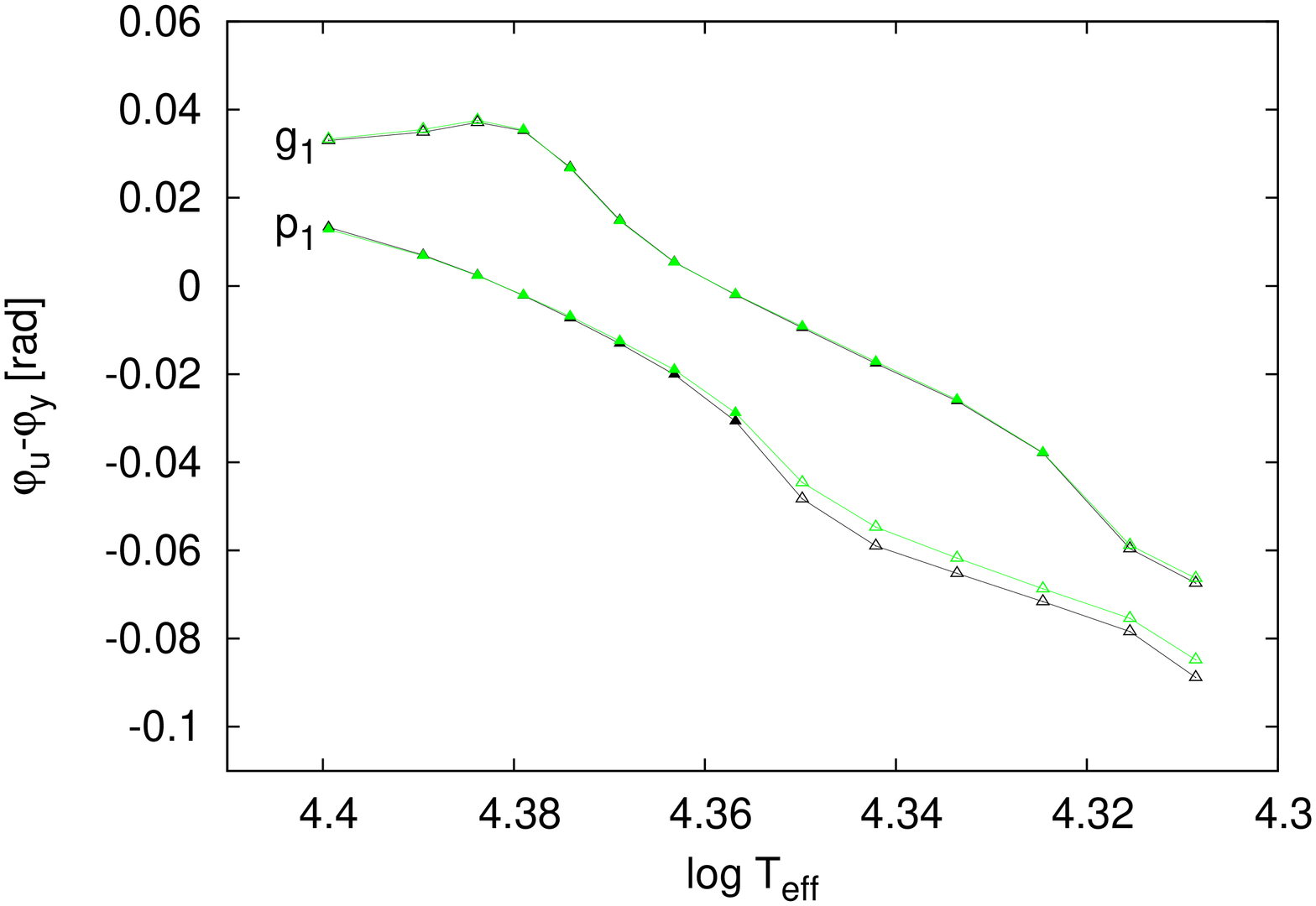}
\caption{The same as in Fig.\,\ref{atm_amp_NLTEl0} but for the two $\ell$=2 modes: p$_1$ and g$_1$. Note different scale on the Y axis.}
\label{aaa}
\end{figure*}
In Fig.\,\ref{aaaa} and \ref{aaa}, we show effects of NLTE on photometric observable
for nonradial modes with $\ell$=1 and 2, respectively.

\section{Conclusions}
The photometric amplitude ratios and phase differences of the early B-type main sequence pulsators strongly depend
on chemical composition and opacities.
The effects of the atmospheric parameters are smaller but may become important
when the nonadiabatic parameter $f$ is determined from observations instead taken
from the pulsation theory.
Here, we studied for the first time the NLTE effects on photometric observables
of the $\beta$ Cep star model.

\acknowledgements
The work has been supported by HELAS EU Network, FP6, No. 026138.


\end{document}